%% file: bergine.tex
\documentclass[multphys,vecphys]{svmult}

\usepackage{makeidx}     
\usepackage{graphicx}    
\usepackage{multicol}    

\makeindex             

\def\CeiO{C$^{18}$O}           
\def\NtwoHp{N$_2$H$^+$}        

\begin{document}

\title*{Astrochemistry and Observations}
\author{Edwin A. Bergin\inst{1}}
\institute{University of Michigan, Ann Arbor, MI 48109 (USA)
\texttt{ebergin@umich.edu}
}
%
%
\maketitle

\section{Introduction}
\label{intro}

The past few years has seen dramatic improvements in our ability to 
extract chemical information from molecular line observations. 
These advances have been driven by new observational platforms and techniques.
In the following I will outline the various methods and uncertainties  
in computing chemical information from observations across the
spectrum of star formation activity (from diffuse clouds, to star-forming
cores, to molecular outflows).
I will show how improved chemical knowledge can be used to place
stronger constraints on theoretical astrochemical models, but also
open new avenues in the use of molecular emission to examine 
the physics of molecular clouds and star formation.

\section{Diffuse Cloud Chemistry}
\label{intro}

Diffuse clouds are transition objects between the low density atomic medium
and the higher density molecular phase.  
The allure of diffuse clouds is twofold.   First they 
are generally, but not exclusively, studied
via absorption lines.
For absorption lines 
the conversion of equivalent width to column density 
is fairly straightforward and requires little information regarding source 
structure.
In contrast, knowledge of the density and temperature structure is often
needed to obtain column densities from emission lines.
Thus chemical abundances derived from absorption lines 
can be considered more reliable.
Second, these objects present
rather simple laboratories to study basic physical and chemical processes.

\begin{figure}[t]
\centering
\includegraphics[height=6cm]{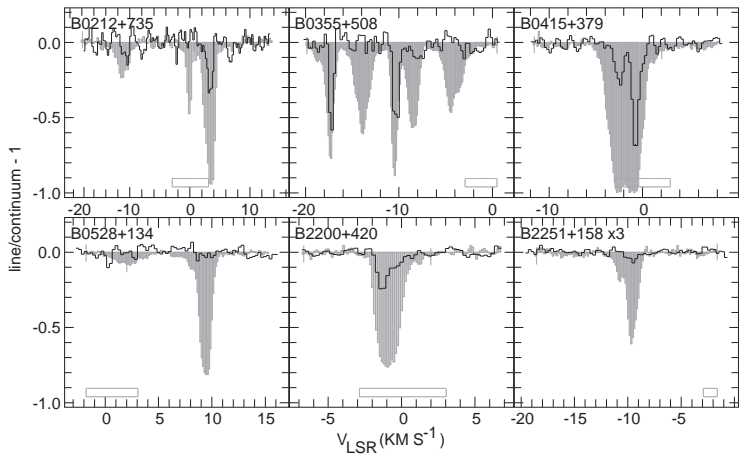}
\caption{CS J=2$-$1 (solid line) and HCO$^{+}$ J=1$-$0 (grey overlay) absorption
spectra towards 6 quasars.  Taken from Lucas \& Liszt (2002).}
\label{diffuse}       
\end{figure}

Over the past decade studies of diffuse cloud chemistry through
absorption lines has 
received significant attention in a series of papers by Lucas and Liszt
(see Liszt \& Lucas 2002 and references therein).   In Figure~\ref{diffuse}
I provide a sample of this work showing detections of HCO$^{+}$ (J=1$-$0)
and CS (J=2$-$1) in absorption in several line of sight clouds.
To study basic chemical 
processes these investigations are able to observe the various species 
that constitute
simple self-contained chemical networks.  For instance, the network that creates
CO in diffuse gas is believed to be:
C$^{+}$ $+$ OH $\rightarrow$ HCO$^{+}$, followed by
HCO$^+$ $+$ e $\rightarrow$ CO $+$ H.
Lucas \& Liszt (2000) find that the column density of HCO$^+$ is capable of
accounting for the observed column density of CO.  However, the measured abundances of
C$^+$ and OH are not high enough to account for observed HCO$^+$.  
Thus either our understanding of ion-molecule gas phase chemistry is incorrect
or diffuse clouds are not just ``simple'' laboratories.  Indeed other more exotic non-thermal
chemical processes have been invoked to overcome these failings (Zsargo \& Federman 2003;
Flower \& Pineau des For\^ets 1998).

\section{Chemistry and Star Formation}

The most significant recent advance in astrochemical studies of
star-forming cores is the development 
of new techniques that move beyond
line of sight average abundances, towards an 
examination of chemical structure as a function of depth.   
This is possible via new instrument
capabilities which have provided images of the dust continuum
emission, or information regarding the spatial 
distribution of dust absorption.
This information has been gleaned through a variety of methods extending from
1.3mm and sub-mm continuum emission images (low-mass star-forming
cores: Andr\'e et al. 2000, Shirley et al. 2001; high-mass
star-forming cores: van der Tak et al. 2000, Hatchell and van der Tak 2003),
to images of dust absorption against the galactic mid-IR background
(Bacmann et al. 2000), and near-IR extinction mapping 
(Alves, Lada, \& Lada 2001).  

Since the dust column density and mass is correlated
with the H$_2$ column density and mass (Hildebrand 1983; Gordon 1995),
these observations provide the clearest information
to date on the column and volume density distribution of H$_2$ molecules.  
This significantly aids the molecular observations in two ways.
(1) By indirectly confirming the location of the H$_2$ 
density and column density peak
and, (2) by providing the density profile which helps to unravel the
similar effects of density and abundance on excitation.

\begin{figure}
\includegraphics[height=3.7cm]{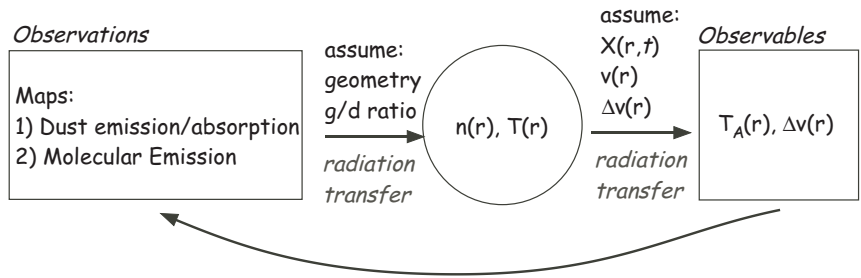}
\caption{ Flow chart describing the 
iterative procedure used to determine abundance structure
with depth.  The abundance profile can either be assumed or predicted 
via a chemical model, in which case it can be time dependent.}
\label{flow}       
\end{figure}

In Figure 2 we present a flow chart that outlines this
new iterative method to derive abundances.  A key to this process is 
that spatial information is important
for observations of both dust and gas.  The first phase involves
using dust observations to constrain core physical properties, typically
the radial density profile.\footnote{For cores with embedded stars 
the dust spectral
energy distribution is used to estimate the radial temperature profile.}
At this point there are 2 methods that can be used to examine the 
abundance structure; both adopting spherical geometry.

(1) In the first method the density and temperature profile\footnote{Along with assumed profiles for radial velocity gradient and
velocity dispersion.}, and an 
assumed abundance profile 
are used as inputs to a radiative transfer model.
The radiation transfer codes adopt either accelerated lambda
iteration or Monte-Carlo methods.  The results from this process are
predictions of the emission spectra as a function of position, which can 
be compared directly to observations.  
An iterative procedure then determines the abundance profile
(see van der Tak et al. 2000; Schreyer et al 2002;
Tafalla et al 2002, 2003; Caselli et al 2002;
Jorgensen et al. 2002; di Francesco et al. 2003; and Lee et al 2003).
(2) The second method uses a chemical model
(using the same density profile 
for the chemical and radiative transfer models)
to predict the abundance profile.
The abundance profile is then used as input to the radiative transfer
producing observable quantities,
which can be iterated by changing 
the parameters in the chemical model or by looking at the time-dependence
(Bergin et al. 2002; Doty et al. 2002).  

In the following we briefly discuss examples for pre-stellar cores
and for sources which contain embedded sources.

\subsection{Chemistry in Pre-Stellar Cores}

\begin{figure}[t]
\includegraphics[height=10.5cm, angle=-90]{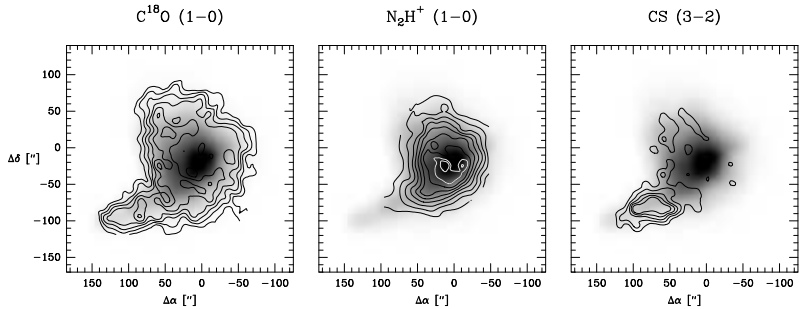}
\caption{ 
The observed emission distribution of visual extinction
or total column density (gray scale) in the B68 dark cloud.  Contours are
integrated emission maps of {\rm C$^{18}$O} (left-hand panel), {\rm N$_2$H$^+$}
(middle), and {\rm CS} (right-hand panel).   The {\rm N$_2$H$^+$} contours
are reversed in color near the central peak to show the distribution
in the center over the extinction distribution.
The emission distributions
illustrate sequence of increasing molecular depletion from {\rm CS}, to {\rm C$^{18}
$O},
and even {\rm N$_2$H$^+$} (Bergin et al. 2002; Lada et al. 2003). 
}
\label{fig-double}
\end{figure}

A representative pre-stellar core is the dark cloud B68
where the centrally concentrated H$_2$ distribution is 
traced through a map of dust visual extinction 
(Alves, Lada, and Lada 2001).   In Figure~3 we present the B68
A$_V$ map along with a series of
molecular emission maps.  In this core
we see that the N$_2$H$^+$ emission peaks in a shell partially
surrounding the peak of dust extinction.
Moreover, the N$_2$H$^+$ peaks inside the much larger C$^{18}$O emission hole, 
which itself lies inside the CS emission depression.

Figure 4 presents the analysis of the N$_2$H$^+$ and C$^{18}$O emission,
which used a gas-grain chemical model linked to a Monte-Carlo radiative
transfer code (Bergin et al. 2002).  
If the emission is in LTE and optically thin then the integrated emission
from a source with constant abundance would appear as a straight
line in this diagram.
Clearly the C$^{18}$O-A$_V$ relation 
deviates from a straight line at
high extinctions and \NtwoHp\ deviates at both high and low A$_V$.  
Such structure in this relation could be due to
freeze-out/depletion or high opacities
(observations of a lesser abundant
isotope confirm freeze-out for CO).
The various solid and dashed lines show emission profiles predicted by
the chemical model for various times.  As time proceeds the \CeiO\ emission
declines via freeze-out; in contrast the \NtwoHp\ emission increases
(as a result of CO freeze-out).
The dotted line is the ``best fit'' radial abundance profile 
which reproduces the emission profile shown as a solid line.
For \CeiO\ the observations require significant freeze-out and also an
abundance reduction at low extinction due to photo-dissociation.  In the
case of \NtwoHp\ an abundance reduction at low extinction is
also required.\footnote{This is due to freeze-out of N$_2$.}

\begin{figure}[t]
\includegraphics[height=3.8cm]{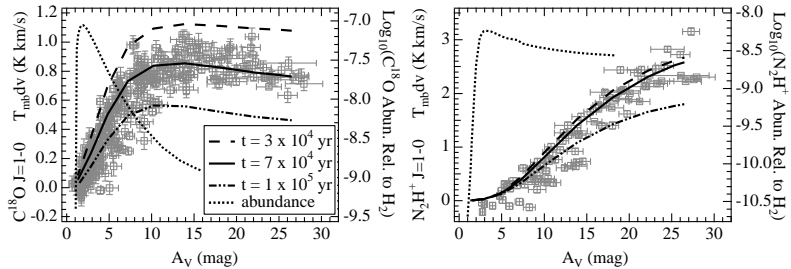}
\caption{ 
Point by point comparison of {\rm C$^{18}$O} 
and {\rm N$_2$H$^+$}  J$=1-0$ integrated intensity
with A$_V$ the entire B68 dark cloud.
In all plots the data are presented as open
squares with error bars while solid curves represent the emission predicted by
a model combining chemistry with a Monte-Carlo radiative transfer code.
The various lines delineate the time-dependence in the chemistry.
The dashed lines are the best fit molecular abundance profiles with the
axis labeled to the right (abundance of {\rm N$_2$H$^+$} is multiplied
by 100).  
Adapted from Bergin et al. (2002).
}
\label{fig-chem}
\end{figure}

B68 is just one example of a core where, using new
techniques, we are inferring
significant freeze-out in the early stages of star formation.  
Other examples can be found in the work
of Tafalla et al. (2002) and references in \S 3.
One possibility that cannot be discounted at present is that nearly
all typically used molecular tracers are absent from the gas in the innermost
regions of these cores (Bergin et al. 2002); this  may hinder
our ability to study the earliest stages star formation. 
However, these sources still offer great promise as new and more
precise laboratories for testing chemical theory than 
previous work in template sources such as TMC-1 and L134N.

\subsection{Chemistry in Star-Forming Cores}

The chemistry in star-forming cores overlaps with the chemical
studies of hot cores.   Hot cores are molecular condensations
directly associated with young embedded massive stars.
Stellar birth leads to the creation of sharp temperature gradients
and higher temperatures lead to the release of frozen ices from grain
surfaces.
As a result hot cores  have quite different chemical
composition, extending to very complex molecular species,
than typically observed in quiescent gas.
While these regions are associated with massive
stars it is clear that low-mass stars also can produce similar
``warm'' cores
(Cazaux et al. 2003). 
The reader is referred to the following 
references for additional discussions 
(Millar 2001; van der Tak 2003). 

Examples of the new iterative method to derive abundances use 
so-called ``jump'' models where abundances are enhanced in
the inner core via grain surface evaporation and are lower in the
outer layers.  The jump transition is set by the evaporation temperature
(see van der Tak 2003).  
Models where chemical networks
are directly linked to the radiation transport code are exemplified by
the work of  Doty et al. (2002) and Boonman et al. (2003).

A generic result is that enhanced abundances of numerous 
molecules are required in the center.  A key goal for hot core
studies is to probe the extent of grain surface chemistry via
observations of trace complex species in the gas.  Grain 
mantles can be observed directly via vibrational modes absorbing
background IR radiation, but this method lacks sensitivity for all but the
most abundant ices.  Thus, gas-phase observations
can be used to probe the extent of grain
surface chemistry beyond the production of 
H$_2$O, CH$_4$, and NH$_3$. 
The problem is that when molecules are released from the grain
they might react in the gas 
(Charnley, Tielens, \& Millar 1992).  
To draw upon a parallel to cometary observations, the products
of these reactions
would be  ion-molecule daughter products as opposed to 
parent molecules that have directly evaporated from the grain surface.
This is further complicated by the potential presence of high temperature
gas-phase chemistry.  Advances in our understanding
will require continued close coupling of chemical models to observations. 

\section{Shock Chemistry}

Another area of observational astrochemistry is the evidence for shock
chemistry detected in molecular outflows.  The primary example is
found in the highly collimated L1157 molecular outflow, 
which has been subjected to
a detailed chemical survey (Bachiller et al 2001).
This study has shown that some species have enhanced abundances in 
outflow lobes (SiO, CH$_3$OH, HCN) while others (N$_2$H$^+$, DCO$^{+}$) are
absent from the flow.  
Another collimated outflow, BHR71, has recently been the focus of
an extensive chemical survey (shown in Figure~5).  In this source
the abundance trends are quite similar to L1157 in regards to
the presence or absence of enhancements for most species.
However, clear differences are found in
the magnitude of the abundance enhancements (Bourke et al. 2004). 
It is not clear whether the differences are intrinsic or due to 
assumptions in the analysis.   

\begin{figure}
\includegraphics[height=5.5cm]{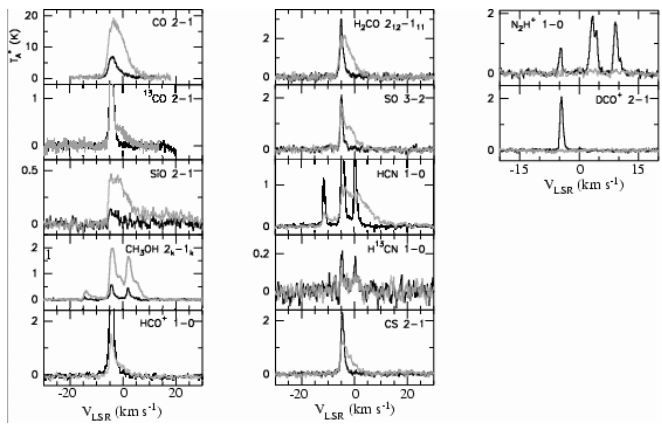}
\caption{ 
Spectra of selected species taken from a chemical survey 
of the BHR71 outflow (Bourke et al. 2004).  Dark lines indicate
spectra taken towards the quiescent core while grey lines are
spectra towards the peak of the red lobe of the outflow.
}
\label{bhr71}       
\end{figure}

Abundance enhancements are typically attributed to grain mantle core and
mantle release via sputtering and/or high temperature chemistry.
The effects of the latter have not been observationally explored
and more attention needs to be paid to models in the literature (see  
series by Pineau des For\^ets and collaborators).  Moreover if shocks
are truly liberating mantles then outflows should show high levels
of deuteration as seen in hot cores.  This also needs to be explored.

\section{Conclusions}

It is clear we are at the beginnings of a new era in astrochemistry
where we will obtain more precise information regarding the abundances
of molecular species, but also set limits on the abundance profile and 
on basic theory.
In Figure~6 we present a schematic that shows a rough
outline of our current understanding of the chemical structure in 
star forming cores.  An important point here is that these zones will
be different for various species (depending on the evaporation temperature,
photo-dissociation cross-sections, depletion rate, etc.).  Since molecular
emission is the only way to probe gas kinematics, one can use
improved chemical information as a tool.  Thus,
with careful choice of tracers, and knowledge of the chemistry,
one can selectively probe motions from different gas
along a given line of sight.  
This will be an important key to unraveling the star formation puzzle.
In all, new observations are leading to crucial tests of chemical theory,
but also presenting a completely new approach to star formation studies,
bringing astrochemistry into wider application as a potent tool.

\begin{figure}
\includegraphics[height=6.0cm]{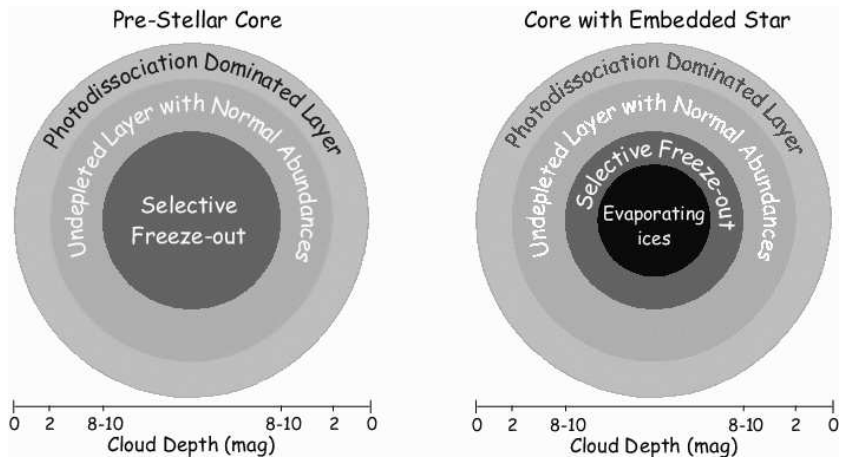}
\caption{ 
Schematic of inferred chemical structure in star-forming cores.
Extinction scales are calibrated to low-mass cores.
}
\label{schematic}       
\end{figure}

%
%
%
\input{bergine_ref}



\printindex
\end{document}

%% file: bergine_ref.tex
%
%

%
%